# Review on Requirements Modeling and Analysis for Self-Adaptive Systems: A Ten-Year Perspective

Zhuoqun Yang [a], Zhi Li [b], Zhi Jin [c,*], He Zhang [d]

[a] *Institute of Mathematics, Academy of Mathematics and Systems Science, Chinese Academy of Sciences, Haidian Dstr., Beijing 100190, P. R. China*
[b] *Software Engineering Dept., College of Computer Science and Information Technology, Guangxi Normal University, Guilin, Guangxi 541004, P. R. China*
[c] *Key Laboratory of High Confidence Software Technologies (MoE), Peking University, Haidian Dstr., Beijing 100871, P. R. China*
[d] *State Key Laboratory of Novel Software Technology, Software Institute, Nanjing University, Hankou Road, Nanjing, Jiangsu 210093 P. R. China*

ABSTRACT

*Context:* Over the last decade, software researchers and engineers have developed a vast body of methodologies and technologies in requirements engineering for self-adaptive systems. Although existing studies have explored various aspects of this field, no systematic study has been performed on summarizing modeling methods and corresponding requirements activities.
*Objective*: This study summarizes the state-of-the-art research trends, details the modeling methods and corresponding requirements activities, identifies relevant quality attributes and application domains and assesses the quality of each study.
*Method:* We perform a systematic literature review underpinned by a rigorously established and reviewed protocol. To ensure the quality of the study, we choose 21 highly regarded publication venues and 8 popular digital libraries. In addition, we apply text mining to derive search strings and use Kappa coefficient to mitigate researchers' disagreements.
*Results:* We selected 109 papers during the period of 2003-2013 and presented the research distributions over various kinds of factors. We extracted 29 modeling methods which are classified into 8 categories and identified 14 requirements activities which are classified into 4 requirements timelines. We captured 8 concerned software quality attributes based on the ISO 9126 standard and 12 application domains.
*Conclusion:* The frequency of application of modeling methods varies greatly. Enterprise models were more widely used while behavior models were more rigorously evaluated. Requirements-driven runtime adaptation was the most frequently studied requirements activity. Activities at runtime were conveyed with more details. Finally, we draw other conclusions by discussing how well modeling dimensions were considered in these modeling methods and how well assurance dimensions were conveyed in requirements activities.

*Keywords:*
Requirements engineering, Requirements modeling, Requirements analysis, Self-adaptive systems, Autonomic computing, Systematic literature review,

## Contents

*E-mail addresses:* zhuoqun.y@gmail.com (Z. Yang), zhili@gxnu.edu.cn (Z. Li), zhijin@pku.edu.cn (Z. Jin), he@software.nju.edu.cn (H. Zhang).

# 1. Introduction

Self-adaptive systems (SASs) are assumed to be capable of adjusting their behaviors in response to meaningful changes in the environment and itself in order to keep continuous satisfying both functional requirements (FR) and non-functional requirements (NFR). The adaptability is characterized by self-configuration, self-healing, self-protection and self-optimization, which are known as self-* properties [1]. Building adaptation mechanism for endowing SASs with the ability of adaptation becomes the main challenge. When developing SASs, engineers should take not only domain logic but adaptation logic into account.

Requirements engineering (RE) comes as the first stage in the development of software systems, aiming at defining domain logic, identifying stakeholders and their needs, and documenting them for subsequent analysis and implementation [2]. Like RE for domain logic, RE for adaptation logic needs to provide answers to the questions similar to the following (5W1H), i.e., what the adaptation requirements are, when and where the system needs to adjust themselves, who is responsible for achieving the adaptation, why the adaptation needs to be performed and how to achieve the adaptation.

During the RE process, requirements analysts should conduct several activities, e.g., requirements elicitation and subsequent requirements modeling and analysis. Requirements modeling and analysis is the fundamental activity in the RE process. It abstractly describes what the target system should achieve by formulating the organization's structures, business rules, goals, tasks, behavior of stakeholders and systems, etc.

This activity is even more important in developing SASs as models produced by requirements modeling and analysis will not only serve as the bases which are amenable to the following requirements activities but also serve as runtime models as the enablers of self-adaptation [3]. For example, building the traceability from requirements models to architectural models has shown to be a reasonable way to derive runtime design decisions [4].

Over the last decade, researchers have developed a vast body of work on requirements modeling and analysis for SASs. Existing roadmaps and surveys [5-11] have summarized the methodologies and technologies proposed in RE for self-adaptive systems, provided insights and outlined challenges for research work. However, to the best of our knowledge, no systematic study has been performed on summarizing and evaluating the emerged modeling methods and corresponding requirements activities. Thus, there is no clear view on where the relevant studies have been published, who has been devoted to the research work, what modeling methods, requirements activities, software quality attributes have been studied, what advantages can modeling methods provide and to what extent modeling methods and requirements activities have been elaborated.

The overall objective of this review article is to systematically investigate the research literature of requirements modeling and analysis for self-adaptive systems from 2003 to 2013. We aim to summarize the state-of-the-art research trends, detail the involved modeling methods and requirements activities, identify the relevant quality attributes and application domains, assess the quality of each study and identify research challenges. This review presents significant extensions to our preliminary review of the literature in [12] with 9 structured research questions, wider time span, 8 more primary studies, 2 more quality assessment questions and more detailed results and discussions. To conduct the investigation and report systematic analysis results, we adopt the research guidelines of systematic literature review [13] in the evidence-based software engineering paradigm [14].

The rest of the article is structured as follows. Section 2 presents a brief history and background of RE for SASs, followed by the elaboration of the research methodology and the review protocol in Section 3. The basic results and findings of our investigation on each research question are described in Section 4, followed by further discussions on modeling methods, modeling dimensions, requirements activities and assurance dimensions in Section 5. Section 6 discusses the threats to validity and corresponding countermeasures, followed by comparison with related literature studies in Section 7. Finally, Section 8 concludes the paper and suggests recommendations for future work.

# 2. Background

## 2.1. Infancy

The concept of self-adaptive systems in software engineering (SE) can date back to 1990s, when researchers and practitioners tried to find a new approach to developing software systems to maintain high-quality software, improve efficiency and enhance robustness. The first definition was proposed by Laddaga [15]: "Self-adaptive software evaluates its own behavior and changes behavior when the evaluation indicates that it is not accomplishing what the software is intended to do, or when better functionality or performance is possible." It implies that the self-adaptive software has not only the ability of self-evaluation and self-reconfiguration, but also the ability of self-optimization.

As self-adaptive software came on to the stage, RE also emerged into the spotlight in SE field and has achieved notable advancement since then. Researchers in RE field proposed requirements monitoring as a method of diagnosing requirements violations caused by environmental changes and invalid assumptions [16-21].

In 2000s, autonomic computing was coined by Paul Horn as a solution to meeting the challenge of complexity. In the same year, Irving Wladawsky-Berger introduced self-managing systems, which can be seen as an instance of "intelligent" computer for coping with the growing complexity of operating, managing and integrating computing systems [22]. The essence of autonomic computing has been identified as self-management, which contains four fundamental properties [23]: *Self-configuration*, *Self-healing* [24], *Self-optimization* and *Self-protection* [25]. There are another two features which underpin the above self-* properties: *Self-awareness* [26] and *Context-awareness* [27-28].

## 2.2. Landmark

Inspired by the early studies on requirements monitoring and autonomic computing, the RE community aims to endow systems with self-* properties at the requirements phase. To this end, innovative methodologies and technologies are highly required for working out requirements models and RE activities. In the past few years, three Dagstuhl Seminars on SASs have been held in January, 2008 [29], October, 2010 [30] and December, 2013 [31] respectively, which delivered definitions to newly emerged concepts, summarized the state-of-the-art in different areas and outlined research directions and challenges.

In *Dagstuhl Seminar 08031*, researchers generally defined self-adaptive systems in the roadmap article [6], which describes and discusses the essential concerns of self-adaptation, including *Modeling dimensions* [32], *Requirements*, *Engineering* [8] and *Assurances*. *Dagstuhl Seminar 10431* focused on four other topics, including *Design space* [33], *Processes* [34], *Decentralization of* *control loops* [35] and *Practical runtime V&V* [36], which are documented in the roadmap article [7]. *Dagstuhl Seminar 13511* concentrated on obtaining assurances for self-adaptive software systems, which include four aspects of concerns: *Assurance criteria and scoping*, *Composition and decomposition of assurances*, *Feedback on assurances*, *Perpetual provisioning of assurances at runtime*. Related materials of the seminar can be found in [37].

Our literature review is closely related to the areas and directions proposed above. Modeling methods will present how requirements and systems are modeled in selected studies. Requirements activities are what RE process can provide in different lifecycle phases and timelines. Quality attributes can be used as benchmarks for estimating adaptation approaches. Application domains outline what kinds of systems are discussed in the SASs paradigm. We also provide further discussions on to what extent modeling dimensions are considered in different modeling methods and to what extent assurances of adaptation are conveyed in different requirements activities.

## 3. Research methodology

Evidence-based software engineering (EBSE) aims to improve decision making related to software development and maintenance by integrating current best evidence from research with practical experience and human values by identifying answerable research questions, searching the literature for the best available evidence to the questions, appraising the quality of the evidence, collecting and aggregating available data for answering the identified questions [38]. The core tool of the evidence-based practice is the Systematic Literature Review (SLR), which is a means of specifying research questions, identifying relevant research, selecting primary studies, assessing their qualities, extracting and synthesizing available data and reporting the analysis results [13]. The whole process of SLR is depicted in Fig. 1, including three phases: *Planning review*, *Conducting review* and *Reporting review*. During the planning review phase, a protocol is produced for defining basic review procedures, on which the conducting phase should depend. We brief the review protocol in this section and report the synthesized results in Section 4. Readers can find more details of our protocol in [39].

### 3.1. Research questions

The high-level goal of this literature research is to review the existing research in the literature of requirements modeling and analysis for self-adaptive systems. We proposed 4 answerable and interpretable research questions (RQ) and decomposed them into 12 sub-questions in Table 1.

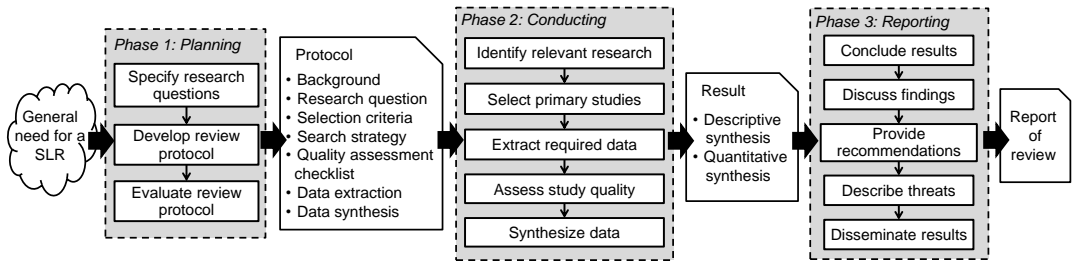

**Fig. 1.** Systematic literature review process.

**Table 1**
Research questions.

| Research Question |
| --- |
| RQ1: What demographic information can we learn from the research publication? |
|     RQ1.1: What is the research distribution over the past decade? |
|     RQ1.2: What is the research distribution over venues? |
|     RQ1.3: What is the research distribution ? |
| RQ2: What primary research concerns can we learn from the selected studies? |
|     RQ2.1: What modeling methods were utilized and what advantages can they provide? |
|     RQ2.2: What requirements activities were studied? |
|     RQ2.3: Which software quality attributes were considered? |
|     RQ2.4: What application domains were used for illustration? |
| RQ3: How well were the selected studies conveyed? |
|     RQ3.1: How rigorously were the modeling methods applied and evaluated? |
|     RQ3.2: How much detail were the requirements activities presented with? |

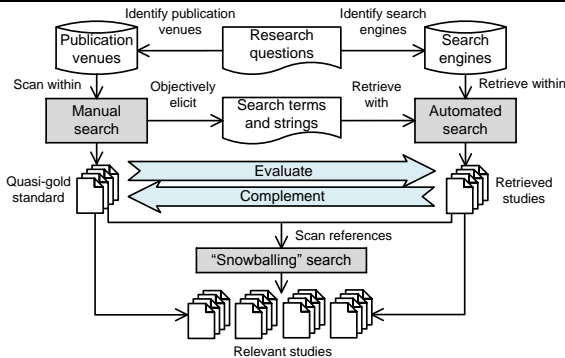

**Fig. 2.** Mechanism underpinning the search process (adapted from [42]).

### 3.2. Search mechanism

The objective of the search process is to identify as many as possible relevant studies according to the research questions. Defining search strategy consists of selecting search method, defining search sources and defining search strings. Appropriate search strategies can optimize the performance of search results and the reliability of the retrieving process. The mechanism underpinning the search process is presented in Fig. 2.

Search sources consist of some publication venues and digital libraries. Publication venues are chosen for conducting manual search, which means researchers scan each identified venues paper by paper and select the relevant ones. Digital libraries are chosen for performing automated search, which means researchers use digital libraries to thoroughly retrieve relevant studies with some search terms and strings adapted to the given search syntax and rules. Manual search is aimed to avoid missing important studies published in domain relevant venues. Automated search is more effective than manual search, but the performance depends on the quality of search strings, capability of digital libraries and diversity of the subject.

According to Zhang and Babar [40], most of the reported SLRs developed the search terms and strings subjectively, e.g., using PICO criteria [13]. Some rigorously-conducted SLRs developed search strings based on the test of collections of "well-known" samples, in order to assess the performance of these strings. Thus, investigators wish to find the gold standard and the most appropriate search strings. However, Kitchenham et al. [41] shows us even though a search process performs well on the chosen gold standard, it is unlikely to be complete on another collection of papers, in other words, determining common gold standard of literature can be hard to implement. Thus, we intend to leverage the *quasi-gold standard* (QGS) [42], which is a set of known studies established by manual search within certain venues and a certain time span, to objectively define search strings and/or evaluate the performance of search strings. In the opposite direction, the retrieved results of automated search complement the results of manual search by expanding the coverage with more relevant studies. "Snowballing" search is another method for expanding the search coverage. It refers to that investigators scan the references in each selected paper derived by manual search and automated search and pick out the most relevant ones as complement. The final set of relevant studies consists of the search results of manual search, automated search and "snowballing" search.

### 3.3. Search sources

Publication venues consist of a collection of proceedings of conferences and symposiums related to the area of RE for self-adaptive systems and major journals and books where the RE community tends to publish their research work. To improve the quality of relevant studies, we choose 11 relevant conferences and symposiums (Table 2) and 10 relevant journals and books (Table 3) according to the Australian Research Council ERA (Excellence in Research for Australia) ranking[1] and H-index[2]. To ensure thorough retrieval, we selected 8 common used digital libraries: IEEE Xplore, ACM Digital Library, Springer, Science Direct, Wiley InterScience, CiteSeerX, EI Compendex and Web of Knowledge.

**Table 2**
Conferences and symposiums.

| Abbr. | Full name | Rank | H-index |
|---|---|---|---|
| ICSE | International Conference on Software Engineering | A | 118 |
| FSE[a] | ACM SIGSOFT Symposium on the Foundation of Software Engineering | A | N/A |
| ASE | International Conference on Automated Software Engineering | A | 44 |
| RE | International Conference on Requirements Engineering | A | 47 |
| REFSQ | International Conference on Requirements Engineering: Foundation for Software Quality | B | 16 |
| MODELS[b] | International Conference on Model Driven Engineering Languages and Systems | B | 22 |
| CAiSE | International Conference on Advanced Information Systems Engineering | B | 44 |
| ICAC | International Conference on Autonomic Computing | B | 32 |
| SASO | International Conference on Self-Adaptive and Self-Organizing Systems | N/A | 9 |
| SEAMS[c] | International Symposium on Software Engineering for Adaptive and Self-Managing Systems | N/A | N/A |
| RE@runtime | International Workshop on Requirements@Run.Time | N/A | N/A |

[a] FSE and European Software Engineering Conference (ESEC) are joint meeting in the odd numbered years.
[b] MODELS has replaced the UML series of Conferences since 2005.
[c] The first SEAMS workshop was held at ICSE 2006. It has become a Symposium co-located to ICSE since 2011.

**Table 3**
Journals and books.

| Abbr. | Name | Rank | H-index |
|---|---|---|---|
| TSE | IEEE Transactions on Software Engineering | A* | 100 |
| TOSEM | ACM Transactions on Software Engineering and Methodology | A* | 49 |
| IST | Information and Software Technology | A | 48 |
| JSS | Journal of Systems and Software | A | 54 |
| ESE | Empirical Software Engineering | A | 33 |
| ASEJ | Automated Software Engineering | A | 26 |
| REJ | Requirements Engineering Journal | A | 27 |
| SoSyM | Software and Systems Modeling | B | 24 |
| TAAS | ACM Transactions on Autonomous and Adaptive Systems | B | 18 |
| SESAS[a] | Software Engineering for Self-Aaptive Systems | N/A | N/A |

[a] SESAS is the outcome of Dagstuhl Seminar on "Software Engineering for Self-Adaptive Systems".

### 3.4. Selection strategy

This subsection introduces the selection criteria, how to use the criteria, how to deal with duplicated papers and repeated studies, and the selection procedure.

#### 3.4.1. Inclusion and exclusion criteria

The set of inclusion criteria and exclusion criteria are provided in Table 4.

The principles of using the inclusion/exclusion criteria are as follows.
- Relevant papers are firstly selected by using inclusion criteria. If one paper meets all the inclusion criteria, i.e., C1 AND C2 AND C3 AND C4, it will be included.
- The remaining papers are selected by using exclusion criteria. If one paper meets any one of the exclusion criteria, i.e., C5 OR C6 OR C7 OR C8, it will be excluded.

**Table 4**
Selection criteria.

| Inclusion criteria |
|---|
| C1: Publication date between 2003.1-2013.12 |
| C2: peer-reviewed conference papers and journal articles |
| C3: Involve concrete modeling methods and requirements activities |
| C4: Involve concrete illustrations for the proposed approaches |
| **Exclusion criteria** |
| C5: Publication in non-English languages |
| C6: In the form of abstract, keynote, poster and short paper (less than 6 pages) |
| C7: Opinion pieces and position papers (without details of modeling methods) |
| C8: Secondary studies, e.g., roadmap, review and survey. |

#### 3.4.2. Duplicated papers and repeated studies

*Duplicated papers.* A duplicated paper refers to a paper which can be retrieved from more than one digital library. This phenomenon is caused by the many-to-many relationships between digital libraries and venues. We should make sure there are no duplicated papers existing in the final collection of relevant studies, so as to guarantee the validity of the SLR. In this situation, we just retain only one of these duplicates in the final set of relevant studies and remove the duplication permanently.

*Repeated studies.* A repeated study refers to the same study published in more than one venue with the similar authorship. It has little contribution to the original research. In this situation, we remove the repeated papers temporarily and retain the most comprehensive or the most recent one. However, we still keep records of them in the reference for reporting the full coverage of the relevant publications.

#### 3.4.3. Selection procedure

We applied the selection criteria at manual search stage, automated search stage and "snowballing" search stage, for establishing QGS, selecting retrieved studies and supplementing the retrieved results respectively. The selection process is performed by two PhD students independently and consists of three rounds in each stage:

**Round 1:** Scaned one paper by title, aiming to eliminate any irrelevant papers. Any paper that any PhD student thinks should be included or unsure about should remain in the set of candidate papers for round 2.

---

[1] Excellence in Research for Australia ranking (2010): http://lamp.infosys.deakin.edu.au/era/?page=cnamesel10.
[2] The H-index for conferences is computed with Microsoft Academy Search: http://academic.research.microsoft.com/, while the H-index for journals is based on SCImago Journal Rank: http://www.scimagojr.com/.

**Round 2:** Read the abstracts of candidate papers from round 1 and appraised each paper with selection criteria. Similar to round 1, any paper that any PhD student considers should be included or unsure about should remain in the set of candidate papers for round 3.

**Round 3:** Looked through the full-texts of the candidate papers from round 2 and assessed each paper with the selection criteria. On one hand, any paper that both PhD students think should be included remained in the set of relevant studies. On the other hand, any paper on which PhD students could not reach agreement should be resolved either by discussion or by arbitration by the supervisors.

Besides, during data extraction, we assessed each paper from round 3 with exclusion criteria. Any paper that meets one of the criteria should be excluded from primary studies. It can be viewed as round 4.

Before round 1, all the duplicated papers should be removed. After round 3, the repeated studies should be removed temporarily. Two supervisors performed a random check (20% of candidate papers) after each of the three rounds to ensure the process was carried out rigorously. During each rounds of selection, we calculated the Kappa coefficient [43] using the algorithm in [44] to certify the inter-rater agreement between the two investigators and minimize possible systematic misunderstandings when making selection decisions. If the Kappa value is above 0.8, it means the understandings of the two investigators are almost consistent.

### 3.5. Manual search and search strings

This subsection presents the establishment of QGS based on the manual search results and the definition of search terms and strings using text mining.

#### 3.5.1. Manual search results

Manual search was conducted by two PhD students individually and was terminated when the Kappa value was larger than 0.8, which denotes a very good agreement [45]. We scanned all papers in the publication venues in Table 2 and Table 3 by title, abstract and full-text with the selection criteria. The Kappa value was above 0.8 and any disagreement was eliminated by discussion. Finally, we selected 62 relevant papers as the QGS in the end of the manual search. The frequency of relevant papers in each publication venue is presented in Table 5. According to the identified venues and digital libraries, the QGS consists of 29 papers from IEEE Xplore, 11 papers from ACM DL, 18 papers from Springer and 4 papers from ScienceDirect.

**Table 5**
Frequency of papers in selected publication venues.

| Conference & symposium | # | % | Journal & book | # | % |
|---|---|---|---|---|---|
| SEAMS | 13 | 26% | REJ | 3 | 25% |
| RE | 6 | 12% | JSS | 3 | 25% |
| RE@runtime | 5 | 10% | SESAS | 2 | 17% |
| ICSE | 5 | 10% | IST | 1 | 8% |
| MODELS | 5 | 10% | TSE | 1 | 8% |
| REFSQ | 4 | 8% | ToSEM | 1 | 8% |
| ASE | 3 | 6% | ASEJ | 1 | 8% |
| CAiSE | 3 | 6% | ESE | 0 | 0% |
| FSE | 3 | 6% | SoSyM | 0 | 0% |
| ICAC | 2 | 4% | TAAS | 0 | 0% |
| SASO | 1 | 2% | — | — | — |
| Total | 50 | 100% | Total | 12 | 100% |

#### 3.5.2. Search terms and strings

Search terms can be elicited based on the QGS in an objective way. One method for eliciting the search terms is using text mining. A frequency analysis of papers' information was undertaken followed by a statistical analysis of most frequently occurring words or phrases. QDA Miner and WordStat[3] are developed textual analysis tools that cannot only determine the most frequent terms but reveal the underlying relationship among these terms.

We imported the title-abstract-keyword segments of each paper in to QDA Miner and chose the *term frequency* (TF) and *inverse document frequency* (IDF) from the algorithms by invoking WordStat. The Jaccard's similarity coefficient enabled us to determine the importance of terms or phrases by comparing the similarity and diversity of the segments of papers.

We derived the clusters of terms according to the frequency and Jaccard's coefficient (Fig. 3). However, though some words have low frequency, they are closely related to requirements activities, e.g., requirements verification and requirements diagnosing. To expand the coverage of the retrieved results, we added these terms and phrases to the set of terms elicited above as complements. Besides, considering synonyms of the elicited terms, e.g., model and modeling, we also added these synonyms into the search strings. The final search strings (Appendix A) are clustered according to requirements activities.

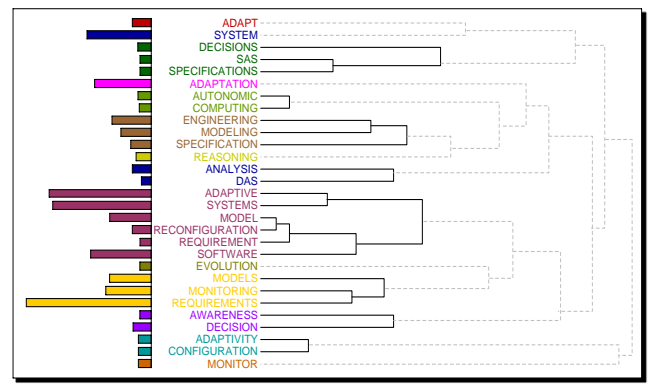

**Fig. 3.** Clustering result of high frequency terms.

### 3.6. Automated search and evaluation

Two PhD students conducted automated search within the identified digital libraries by splitting and inputting the strings according to the search syntax and search capability of each engine. To assess the feasibility of the defined automated search process, two supervisors conducted the pilot study with a subset of digital libraries. After eliminating disagreement between investigators, we selected 105 papers at the automated search stage.

*Quasi-sensitivity* is an important criterion for evaluating the quality and efficiency of a search strategy [42]. It refers to the proportion of relevant studies covered by QGS and can be calculated as:

$$quasi\text{-}sensitivity = \frac{number\ of\ retrieved\ relevant\ studies\ covered\ by\ QGS}{total\ number\ of\ relevant\ studies\ in\ QGS} \cdot 100\%$$

According to the venues in Table 5, quasi-sensitivity of search strings should be computed separately with the selected results of IEEE Xplore, ACM DL, Springer and ScienceDirect and then a total quasi-sensitivity should also be calculated. Fig. 4 provides the quasi-sensitivity for the selected results of each digital library and the total quasi-sensitivity for the whole QGS. According to [42], once the sensitivity is above 80%, the search strategy can be considered optimum. The search strings perform well for IEEE Xplore, ACM DL, Springer and ScienceDirect with the sensitivity of 96.55%, 100.00%, 88.89% and 100.00% respectively. Meanwhile, the total quasi-sensitivity is 95.16%. Thus, the search strings and search strategy can be considered optimum in our SLR.

---
[3] QDA Miner 4 and WordStat 6: http://provalisresearch.com/products/.

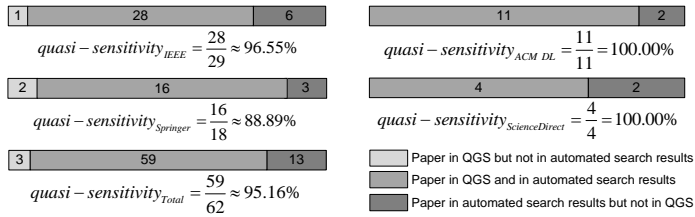

**Fig. 4.** Quasi-sensitivity for each digital library and the total quasi-sensitivity.

### 3.7. Search and selection results

Fig. 5 elaborates the detailed selection process and results as the final primary studies. The left part presents the removed results while the right part shows the remaining results. The total number of removed papers is 2388. The final primary studies are provided in Appendix B.

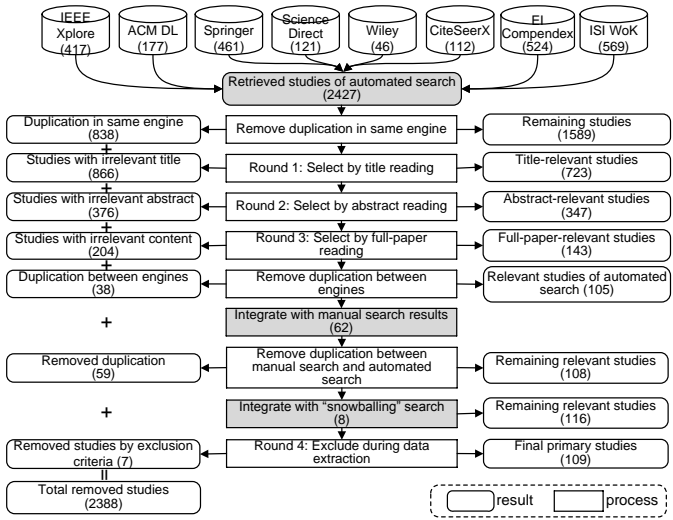

**Fig. 5.** Details of the search and selection process.

### 3.8. Quality assessment

To answer RQ3, we need to assess the quality of each primary study. The quality assessment checklist (Appendix C) is designed based on the assessment items introduced in [13, 46]. The checklist is not used to evaluate an individual method, but for evaluating how the methods and requirements activities are conveyed in the literature. The assessment process was conducted by the two PhD students individually and the final scores were calculated by integrating their evaluation results.

### 3.9. Data extraction

To answer the research questions in Table 1, a data extraction form (refer to our protocol [39]) was designed to record the relevant data that can be extracted from the selected primary studies. The extraction process was also conducted by two PhD students. Two supervisors perform the pilot study based on a subset of primary studies to check whether the design of extraction process has any systematic flaws. The details of extracted data can be found in the online file [47].

## 4. Results

This section presents answers to the research questions and provides primary findings and discussions by analyzing the extracted data.

### 4.1. RQ1: What demographic information can we learn from the research publication?

By answering RQ1, we present the distributions of publication date, venues, research work, researchers and research groups.

#### 4.1.1. RQ1.1: What is the research distribution over the past decade?

Fig. 6 depicts that though the definition of self-adaptive software was proposed in 1990s, no publication in the literature appeared in 2003 and 2004. Since 2005, the qualified papers in the literature have been steadily increasing in the past decade. The anomalous data in 2013 are caused by a number of short papers, position papers and keynotes, which do not contribute to the qualified publication according to our exclusion criteria. The rapid increase in 2009, 2011 and 2012 may be caused by the Dagstuhl Seminar held in 2008 and 2010. Analogously, inspired by Dagstuhl Seminar 2013, the number of qualified publications will increase in 2014 and 2015 again.

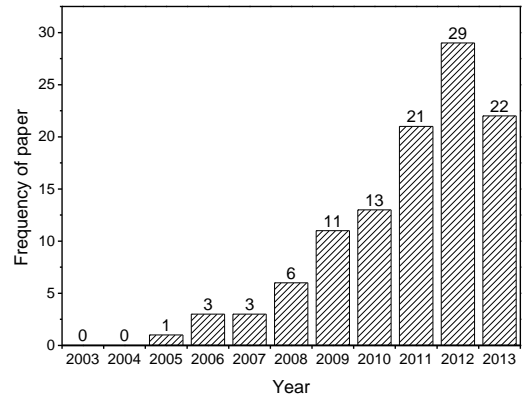

**Fig. 6.** Research distribution over the last decade.

#### 4.1.2. RQ1.2: What is the research distribution over venues?

The distribution of publication over conferences and symposiums is presented in Table 5. SEAMS is the primary venue where the RE community is prone to publish the research work related to self-adaptive systems, followed by RE conference, ICSE and MODELS. RE@runtime was held only in 2010 and 2011, as the workshop in conjunction with the RE conference. The workshop explored the potential for runtime abstractions and models of requirements.

The distribution of publication over journals and books is also shown in Table 5. REJ and JSS involve more primary studies than other journals, while TSE, IST and TAAS involve more secondary studies related to self-adaptive systems. SESAS is the publication of Dagstuhl Seminar, which contains a series of roadmap papers and vision papers.

#### 4.1.3. RQ1.3: What is the research distribution over research groups, countries and continents?

This question is answered by extracting the affiliation information of each selected paper. The selected studies are primarily led by 33 research groups. The distribution of research work (more than 2) over research groups is presented in Fig. 7. Michigan State University contributed to the literature with more publications, followed by Politecnico di Milano and University of Trento. About 82% (27/33) research groups lead less than 4 research work. The top 6 groups in Fig. 7 lead 52% (57/109) of the primary studies.

A total of 16 countries primarily carried out the research work. Fig. 8 depicts that research groups (with more than 2 research work) in Italy, USA, UK, China and Canada contributed 81% (88/109) of the primary studies. Furthermore, Fig. 9 reveals that most of the selected research work is led by European groups (56%), followed by American groups (28%) and Asian groups (16%).

These results provide the research status for researchers to establish cooperative relationships between research groups and countries.

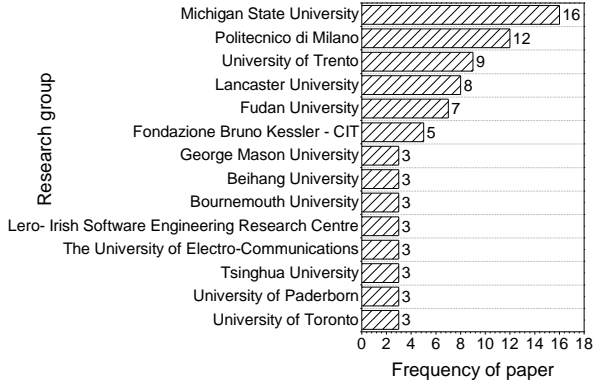

**Fig. 7.** Research distribution over research groups.

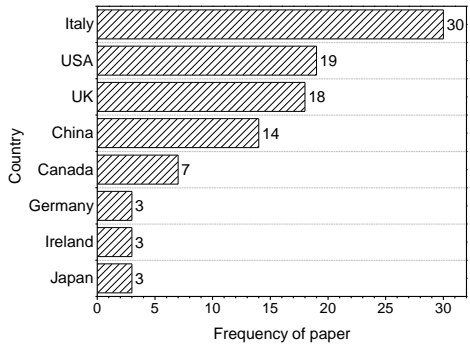

**Fig. 8.** Research distribution over countries.

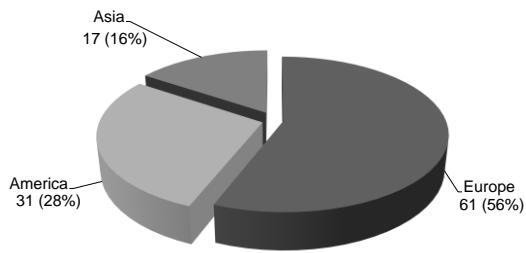

**Fig. 9.** Research distribution over continents.

### 4.2. RQ2: What primary research concerns can we learn from the contents of selected studies?

This subsection presents the used modeling methods, describes advantages of different modeling methods, identifies the involved requirements activities, and summarizes the software quality attributes and the application domains concerned in the literature.

#### 4.2.1. RQ2.1: What modeling methods were utilized and what advantages can they provide?

Requirements models and system behavior models are essential runtime entities for self-adaptive systems. RQ2.1 focuses on the modeling methods for both requirements and system behavior. To answer this question, we adopt and extend the general categories of RE modeling methods provided in [2]. Totally, we extracted 29 modeling methods from the primary studies, 34% (37/109) of which utilized more than one method while 66% (72/109) of which only utilized one method. These modeling methods can be classified into 8 categories. Table 6 provides the categories, frequency, descriptions and optional sub-categories. The most frequently used category is enterprise models, followed by logic model, behavior model, language model, evolutionary computation model, utility model, domain model and network model.

*Enterprise models*

The related enterprise models are presented in Table 7. KAOS, i* and Tropos are goal-oriented methodologies. Among the primary studies that utilize enterprise models, goal-oriented studies account for 83% (55/66). KAOS [18] is the most widely used goal-oriented method in the literature, in which, *goals* model stakeholders' intentions while *tasks* model the FR that can be used to achieve the goals. Goals can be refined with sub-goals or tasks through AND/OR-decompositions. *Softgoals* model the NFR, which have no clear-cut criteria for their satisfaction and can be used to evaluate different choices of alternative tasks. The model elements of KAOS, i.e., goal, task and softgoal, provide avenues for building adaptation mechanisms and achieving adaptation. Based on the partial satisfaction of goals [48, 49], Baresi et al. [S9] proposed the adaptive goal by using fuzzy logic and membership functions to quantify the satisfaction of goals. The AND/OR-decomposition provides rational clues for diagnosing requirements. Wang et al. [S102, S103] provided approaches for monitoring and diagnosing requirements by reasoning on propositional logic. Besides, OR-decompositions also provide optional tasks for realizing a goal. The goal and its tasks can be considered as variation point and candidate configurations of the system. Based on this, Liaskos et al. [S89] proposed an approach for achieving adaptation by selecting these alternatives. Softgoals are related to systems' quality attributes and stakeholders' preferences, which can be used as criteria for selecting adaptation decisions. Sutcliffe and Sawyer [S18] addressed that during adaptation, personal requirements of an individual user should also be concerned. Except for utilizing KAOS model elements, some new elements can be added in for producing richer semantics. For example, awareness requirements in the

**Table 6**
Categories and descriptions of modeling methods.

| Model category (frequency) | Description | Optional models or methods | Abbr. |
|---|---|---|---|
| Enterprise model (66) | Enterprise modeling and analysis deals with understanding an organization's structure, the business rules, goals, tasks and data it needs, generates and manipulates | • KAOS<br>• i*<br>• Tropos<br>• Feature Model<br>• Business Process Model<br>• Problem Frame | KAOS<br>i*<br>Tropos<br>FM<br>BPM<br>PF |
| Behavior model (21) | Behavior modeling describes the dynamic or functional behavior of the system by abstractly modeling the states of the system and their transitions. | • Discrete Time Markov Chain<br>• Label Transition Systems<br>• Finite State Machine | DTMC<br>LTS<br>FSM |
| Logic model (36) | Logical modeling deals with specifying requirements, properties and adaptation logic that need to be satisfied through system behavioral changes. | • Markov Decision Process<br>• Petri Net<br>• UML (activity diagram)<br>• UML (sequence diagram)<br>• UML (statechart)<br>• Linear Temporal Logic<br>• Fuzzy Logic<br>• Probabilistic Computation Tree Logic<br>• Propositional Logic<br>• Computation Tree Logic<br>• Fuzzy Branching Temporal Logic | MDP<br>PN<br>UMLBM<br>ib.<br>ib.<br>LTL<br>FL<br>PCTL<br>PL<br>CTL<br>FBTL |
| Domain model (3) | Domain modeling provides the detailed reasoning about the domain assumptions and the access to requirements reuse within the same domain by explicitly describing the world in which the target system will operate. | • Domain-Specific Model (use case model)<br>• Domain-Specific Model (variability model) | DSM<br>ib. |
| Network model (3) | Network models are used in decision theory to support qualitative and quantitative analysis and decision making by modeling the causal and rational relationships between nodes. | • Dynamic Decision Network<br>• Causal Network | DDN<br>CN |
| Evolutionary computation model (6) | Evolutionary computation models are mathematical models that are used to derive the optimal system configuration. | • Genetic Algorithms<br>• Genetic Programming | GA<br>GP |
| Language models (6) | Language models are produced to specify and formalize system requirements. | • Object Constraint Language<br>• Structured Languages | OCL<br>SL |
| Utility model (4) | Utility model considers that the satisfaction of requirements can be modeled as utility and calculated with utility functions. | • Utility Theory | UT |

KAOS model are proposed for monitoring requirements [S44, S45]. Cheng et al. [S22] modeled environmental uncertainties by integrating the KAOS model with a threat model. By flexibly using KAOS method, both business logic and adaptation logic can be built precisely.

i* method is defined to model and reason about both the system and its organizational environment [50]. There are two basic diagrams of an i* model: the strategic dependency (SD) diagram is used to depict the dependency relationships among various actors while the strategic rationale (SR) diagram presents the internal details of each actor. Actors have intentional properties, such as goals and beliefs. Inside the actor, there are four key model elements: goals, softgoals, tasks and resources. The concepts of goals, softgoals and tasks are similar to corresponding notions of KAOS model. The resources represent data, information, or a physical resource that an actor may provide or receive. There are four dependencies defined based on these four internal elements. A goal dependency means that the depender needs the dependee to achieve a goal. Similarly, a softgoal dependency or a task dependency refers to that the depender needs the dependee to achieve a softgoal or a task. A resource dependency addresses that the depender needs resources provided by the dependee. Tropos method adopts the concepts offered by i* framework. It not only supports early requirements, but later requirements, architecture design and detailed design [51]. According to the results, 77% (10/13) of i* models and 71% (5/7) of Tropos models were used for requirements activities at requirements time, such as building adaptation mechanisms based on dependency relationships [S3, S4] or based on alternatives tasks [S5, S6]. By adding *claim* notations, the requirements model is extended with the ability of self-explanation [S10, S24, S41]. To model self-organizing multi-agent systems, Tropos method models each agent as an individual actor [S19]. Besides, Tropos method can also be used to model requirements and dependencies of socio-technical systems [S59]. Compared with KAOS method, i* and Tropos focus more on the relationship between different actors in organization and the internal logic of these actors.

Feature Model (FM) provides an integrated business view emphasizing on possible variations and changes in its runtime behavior. Feature model is hierarchically structured with a set of features which can be considered as functionalities and variability relationships between features, such as mandatory, optional and alternative [52]. A variant point of a feature model refers to the node on which configuration decisions should be made according to the contextual changes, while a variant is denoted with an optional feature. Self-adaptive systems can achieve a variety of functionalities by binding or removing optional features at runtime [S14, S67, S108]. Feature model focuses on functionalities of a system while goal model emphasizes stakeholders' intentions and ways to achieve them, while compared with goal models, feature model is more close to the architectural level. Both of them can specify variant points with alternative tasks or features. However, the satisfaction of a feature is crisp while the satisfaction of a goal can be assigned with a fuzzy value. Thus, when building adaptation mechanisms, goal-based methods are more flexible, but when considering implementing the reconfigurations, feature-based methods are more direct.

**Table 7**
Enterprise models and related studies.

| Option | Primary study | # | % |
|---|---|---|---|
| KAOS | [S1, S2, S7, S9, S11, S15, S16, S18, S22, S23, S38, S42, S44, S45, S50, S51, S56, S57, S60-S63, S66, S70, S71, S73, S82, S88, S89, S102-S104, S106-S108] | 35 | 53% |
| i* | [S3-S6, S10, S17, S20, S21, S24, S41, S49, S84, S86] | 13 | 20% |
| Tropos | [S19, S25, S40, S43, S48, S55, S59] | 7 | 11% |
| Feature Model | [S14, S58, S67, S72, S76, S80, S81, S108] | 8 | 12% |
| Business Process Model | [S12, S75] | 2 | 3% |
| Problem Frame | [S13] | 1 | 1% |

Business Process Model (BPM) and Problem Frame (PF) are less used enterprise models. Business process is an essential element in development of information systems. Each system or organization may have different ways to perform a process, so achieving flexibility in business processes becomes very important. Santos et al. [S12] proposed a process for supporting the configuration of business processes using non-functional requirements according to contextual changes. The business process execution language (BPEL) has been widely used for web service composition to define business logic via modeling message exchange sequences in an executable manner. Mosincat et al. [S75] introduced how to automatically evolve the system model based on BPEL. The problem frames provide a basis for analyzing software problems and their context. Problem diagrams describe various problem domains, machine, requirements, domain interfaces and requirements references [53]. Salifu et al. [S13] adopted variant frame and variant problem to capture and reason about contextual variability. However, when the problem domain and requirements get more complex, both the context diagram and the problem diagram may lose efficiency of representing the adaptation problem, though it can be divided into several interconnected sub-problems. Besides, there are few reports on how the specification of shared phenomena can be used for reasoning and deriving new problem diagrams. These questions of the problem frame need further research and explanation.

*Behavior models*

The behavior models and related primary studies are provided in Table 8. Discrete Time Markov Chain (DTMC) and Markov Decision Processes (MDP) are Kripke structure with probabilistic transitions among states. States represent possible configurations of the system. Transitions among states are assigned with probability. A MDP can be viewed as the extension to a DTMC with operational actions and rewards [54]. The DTMC is known as a useful formalism to describe systems from the reliability viewpoint and to support reasoning about system properties. These models are involved in the studies of verifying reliability requirements [S95-S98] and maintaining the reliability requirements [S54, S64, S65]. PRISM[4] model checker is an effective tool for modeling and analyzing DTMC and MDP.

Label Transition System (LTS) has been widely used for behavior modeling. It consists of a set of states representing the system configuration and a set of labels between states representing the transition conditions. SASs can perform different behaviors by switching states according to the labels changes. The use of LTSs concentrates on providing general properties, e.g., liveness [S28, S32, S33] and verifying requirements through model checking [S28, S46, S90]. LTSA[5] is a popular model checker for modeling and analyzing LTSs. Compared with DTMC and MDP, the transition conditions of LTS are not probabilistic events.

Finite State Machine (FSM) can be extended with clock variables and used for verifying the requirements and properties specified with Timed Computation Tree Logic [S93, S94, S99]. UPPAAL[6] is a model checker for modeling, validation and verification of real-time systems modeled as networks of timed automata.

Petri Net (PN) is a kind of mathematical modeling language of distributed systems. A Petri Net consists of places, transitions and arcs [55]. Zhang and Cheng [S29] introduced an approach to creating formal models for the behavior of adaptive programs based on Petri Net. Compared with the above behavior models, PN is harder to build because the Token transition is complex for SASs.

Besides, UML also provides diagrams to represent system behaviors. Activity diagrams are used to model workflows [S85]. Sequence diagram are utilized to represent the adaptation scenarios [S8]. Statechart are used to model system behaviors with hierarchically nested states [S61].

**Table 8**
Behavior models and related studies.

| Option | Primary study | # | % |
|---|---|---|---|
| Discrete Time Markov Chain | [S54, S64, S65, S95-S98] | 7 | 33% |
| Label Transition Systems | [S28, S32, S33, S46, S90] | 5 | 23% |
| Finite State Machine | [S77, S93, S94, S99] | 4 | 19% |
| Markov Decision Process | [S85] | 1 | 5% |
| Petri Net | [S29] | 1 | 5% |
| UML (activity diagram) | [S85] | 1 | 5% |
| UML (sequence diagram) | [S8] | 1 | 5% |
| UML (statechart) | [S61] | 1 | 5% |

---

[4] PRISM model checker: http://www.prismmodelchecker.org/.
[5] Labelled Transition System Analyzer (LTSA): http://www.doc.ic.ac.uk/ltsa/.
[6] UPPAAL homepage: http://uppaal.org/.

*Logic models*

Table 9 lists related primary studies on logic models, which are mainly used for activities at requirements time (18/36) and runtime (17/36). Linear Temporal Logic (LTL) is utilized to specify adaptation logic and requirements [S28, S30, S31, S46]. Fuzzy logic (FL) is used for deriving resilient specification [S1, S2] and dealing with uncertainties [S87]. Probabilistic Computation Tree Logic (PCTL) is a specification language for model checking within DTMC and MDP [S95-S97]. Propositional Logic is mainly used to perform reasoning on goal models with the SAT solver [S69, S102, S103]. Computation Tree Logic (CTL) functions as the formal language for specifying properties in timed automata [S93, S94, S99]. Fuzzy Branching Temporal Logic (FBTL) extends branch temporal logic with fuzzy notations for deriving relaxed specifications [S35, S36]. Some details of these formal methods can be found in [10].

**Table 9**
Logic models and related studies.

| Option | Primary study | # | % |
|---|---|---|---|
| Linear Temporal Logic | [S28, S30-S33, S38, S46, S47, S90-S92, S100] | 12 | 33% |
| Fuzzy Logic | [S1, S2, S7, S9, S35-S37, S74, S86, S87, S101] | 11 | 31% |
| Probabilistic Computation Tree Logic | [S54, S95-S97] | 4 | 11% |
| Propositional Logic | [S43, S69, S102, S103] | 4 | 11% |
| Computation Tree Logic | [S93, S94, S99] | 3 | 8% |
| Fuzzy Branching Temporal Logic | [S35, S36] | 2 | 6% |

*Domain models, network models, evolutionary computation models, language models and utility models*

The results of all the rest modeling methods are provided in Table 10. Use cases define the interaction between actors and the target system, model the application logic and represent enterprise architecture. We consider the use case model as a domain-specific model, because application scenarios are captured in the model. *Adapt Cases* extend use cases to allow the explicit description of adaptivity on a higher logical level [S26, S27]. Differently, Bencomo et al. [S52] proposed a domain-specific modeling language based on the variability model, to model the solution architecture and the set of valid system configurations.

Dynamic Decision Network (DDN) is constructed with a set of state features, a set of possible actions forming the decision nodes, and a set of reward functions [56]. It provides a mechanism for making rational decisions. Bencomo et al. [S83] introduced how to make adaptation decisions by combining probability and utility theory together. In [S84], they proposed a decision mechanism by mapping a goal model to a DDN. Causal Network (CN) builds the cause-and-effect relationships between nodes. Salehie et al. [S88] proposed how to measure the effects of threats to personal assets by merging the concept of asset, goal and threat into a CN.

Evolutionary computation involves continuous optimization and combinatorial optimization problems. Genetic Algorithm (GA) and Genetic Programming (GP) are two widely used optimization algorithms. GA and GP are used to derive optimal adaptation decision based on mathematical models [S68, S78, S79, S105], goal models [S82] and feature models [S72].

Object Constraint Language (OCL) is declarative language for describing rules that apply to UML models. Souza et al. [S44, S45] proposed the notion of awareness requirements and introduce the formalization of these requirements based on OCL. Other structural language includes GoalSPEC [S39] which is designed for goal model, RELAX [S34] which is proposed for providing resilience and other languages for defining a system [S53] and formalizing the evolution process [S109].

Utility is an important concept in economics and game theory. It is used to represent the satisfaction experienced by the consumer.

In RE area, utility is adopted to representing the satisfaction of requirements. For example, the satisfaction degree of requirements computed with membership functions can be considered as this type of utility [S9]. In other studies, utility is used as criteria for evaluating the adaptation decisions.

**Table 10**
Domain models, network models, evolutionary computation models, language models, utility models and related studies.

| Option | Primary study | # |
|---|---|---|
| Domain-Specific Model (use case model) | [S26, S27] | 2 |
| Domain-Specific Model (variability model) | [S52] | 1 |
| Dynamic Decision Network | [S83, S84] | 2 |
| Causal Network | [S88] | 1 |
| Genetic Algorithms | [S68, S72, S79, S82, S105,] | 5 |
| Genetic Programming | [S78] | 1 |
| Object Constraint Language | [S44, S45] | 2 |
| Structural Languages | [S34, S39, S53, S109] | 4 |
| Utility Theory | [S82, S83, S104, S105] | 4 |

*4.2.2. RQ2.2: What requirements activities were studied?*

Before answering RQ2.2, we need to define requirements activity timelines. Requirements activity timelines refer to when the requirements activities take place, including *requirements time*, *design time*, *development time* and *runtime*. During the extraction, we totally derive 14 requirements activities which can be classified according to the defined timelines. In this sub-section, we introduce what aspects each activity focuses on. More detailed discussions and research recommendations about modeling methods and requirements activities will be provided in Section 5.

Table 11 depicts that 44% (48/109) of primary studies focus on requirements activities at requirements time. Building requirements-driven adaptation mechanisms puts emphasis on how to build adaptation process based on requirements, while modeling requirements with adaptation logic concentrates on effective representation of requirements. Specifying requirements with adaptation logic refers to expressing requirements with logic models, e.g., LTL [S28, S30] and FBTL [S35, S36], or structural languages, e.g., KAOS specifications [S38] and RELAX [S34]. Building requirements monitoring mechanisms addresses how to effectively monitor requirements, e.g., using awareness requirements [S44, S45] or monitoring algorithms based on goal models [S43]. Verifying requirements and validating requirements for SASs address how to verify requirements based on model checking [S46, S47] and how to detect inconsistency [S48], respectively.

Primary studies that address requirements activities at design time account for 5% (5/109). Mapping from requirements to architecture refers to deriving architectural models based on related requirements models. Designing adaptation mechanisms based on requirements focuses on how requirements model can help design adaptation mechanisms. Validating requirements with design decision aims to test whether the self-adaptation layer can ensure satisfaction of requirements at design stage. Development time also involves 5% (5/109) of primary studies. These studies focus on development by using software product line [S54, S58], development process [S55, S56] and customization [S57].

Forty-seven percent (51/109) of primary studies concentrate on requirements activities at runtime. For this timeline, requirements become runtime entities, which need to be monitored, detected and verified. System adaptation to mitigate requirements violations addresses how to realize adaptation through decision making. Verifying requirements for SASs is concerned with ensuring satisfaction of both FR and NFR at runtime. Monitoring requirements and detecting deviations deal with the problem of how to monitor requirements and diagnose violations. Requirements-driven system evolution concerns achieving evolution based on requirements models at runtime.

*4.2.3. RQ2.3: Which software quality attributes were considered?*

To answer this question, we adopt the software quality attributes identified by ISO 9126 standard. We totally extract 8 software quality attributes. Fifty-eight percent (63/109) of the primary studies focus only on *adaptability*, while 42% (46/109) of studies address other quality attributes, including *changeability*, *time behavior*, *fault tolerance*, *reliability*, *understandability*, *replaceability* and *resource utilization*. The description of each quality attribute and corresponding primary studies are provided in Table 12. These quality attributes can function as criteria for evaluating adaptation mechanisms. If an adaptation mechanism claims lower overhead (e.g., time and resource), higher security, less effort for changing behavior or replacing components and less effort for understanding the adaptation, it can be more efficient and reliable than other mechanisms.

Ninety-four percent (17/18) of the primary studies that concern changeability focus on runtime activities, including 13 in system adaptation, 2 in requirements verification [S95, S96], 1 in system evolution [S107] and 1 in requirements monitoring [S103]. The rest one concerns requirements validation at design time [S53]. Ninety-one percent (11/12) of primary studies that concern time behavior address runtime activities, including 8 in system adaptation, 2 in requirements verification [S95, S96] and 1 in requirements monitoring [S103]. The rest one also concerns requirements validation at design time [S53]. Fault tolerance is mainly involved in activities at requirements time (73%, 8/11), including 5 in building adaptation mechanism [S1, S2, S7-S9] and 3 in specifying requirements with adaptation logic [S34-S36]. Torres et al. [S74] and Ramirez et al. [S86] addressed fault tolerance in runtime adaptation while Iftikhar and Weyns [S94] discussed this attribute in requirements verification at runtime.

Studies involving reliability all focus on runtime activities, including 3 in runtime adaptation [S64, S65, S68] and 3 in runtime verification [S91, S97, S98]. Understandability is discussed in studies which concentrate on requirements time. Bencomo et al. [S10] and Welsh and Sawyer [S24, S41] assured understandability by endowing the system with the ability of self-explanation, while Morandini et al. [S25] discussed how the specified requirements can be understood by users through controlled experiments. Though security is a critical attribute of self-adaptive systems, it lacks discussion. How to protect users' privacy requirements under attack can be found in [S42, S88, S90]. For replaceability, Ma et al. [S6] addressed the replacement of service while Moisan et al. [S67] described a feature-based approach for component replacement. Finally, resource utilization is only discussed in runtime monitoring activity [S105].

**Table 11**
Categories of requirements activities and related studies.

| Timeline | Requirements activity | Phrase for short | Primary study | # | % |
|---|---|---|---|---|---|
| Requirements time | Building requirements-driven adaptation mechanisms | Adaptation mechanism | [S1-S15] | 15 | 14% |
| | modeling requirements with adaptation logic | Modeling req. | [S16-S27] | 12 | 11% |
| | Specifying requirements with adaptation logic | Specifying req. | [S28-S39] | 12 | 11% |
| | Building requirements monitoring mechanism | Monitoring mechanism | [S40-S45] | 6 | 6% |
| | Verifying requirements for SASs | Req. verification | [S46, S47] | 2 | 2% |
| Design time | Validating requirements for SASs | Req. validation | [S48] | 1 | 1% |
| | Mapping from requirements to architecture | Mapping to Arch. | [S49, S50] | 2 | 2% |
| | Designing adaptation mechanisms based on requirements | Mechanism@design-time | [S51, S52] | 2 | 2% |
| | Validating requirements with design decision | Validation@design-time | [S53] | 1 | 1% |
| Development time | Developing SASs from the RE perspective | SASs Development | [S54-S58] | 5 | 5% |
| Runtime | System adaptation to mitigate requirements violations | Adaptation@runtime | [S59-S90] | 32 | 29% |
| | Verifying requirements for SASs | Verification@runtime | [S91-S100] | 10 | 9% |
| | Monitoring requirements and detecting deviations | Monitoring@runtime | [S101-S105] | 5 | 5% |
| | Requirements-driven system evolution | Evolution@runtime | [S106-S109] | 4 | 4% |

**Table 12**
Descriptions of concerned software quality attributes and related studies.

| Quality attribute | Description | Primary study | # | % |
|---|---|---|---|---|
| Adaptability (only) | It is the inherent characteristic of SASs, describing the ability of software to adjust behavior in response to internal or external changes. | [S3-S5, S11-S23, S26-S33, S37-S40, S43-S52, S54-S58, S61, S71, S75, S76, S78, S79, S83, S84, S87, S89, S92, S93, S99-S102, S104, S106, S108, S109] | 63 | 56% |
| Changeability | It characterizes the amount of effort to change a system. | [S53, S59, S60, S62, S63, S66, S69, S70, S72, S73, S77, S80, S81, S85, S95, S96, S103, S107] | 18 | 17% |
| Time behavior | It characterizes response times for a given thru put. | [S53, S62, S63, S66, S72, S73, S80, S82, S85, S95, S96, S103] | 12 | 11% |
| Fault tolerance | It characterizes the ability of software to withstand (and recover) from component or environmental failure. | [S1, S2, S7-S9, S34-S36, S74, S86, S94] | 11 | 10% |
| Reliability | It is broadly defined as the probability of successfully accomplishing an assigned task when it is requested. | [S64, S65, S68, S91, S97, S98] | 6 | 6% |
| Understandability | It determines the ease of which the systems functions can be understood | [S10, S24, S25, S41] | 4 | 4% |
| Security | It is related to unauthorized access to the software functions. | [S42, S88, S90] | 3 | 3% |
| Replaceability | It characterizes how easy it is to exchange a given software component within a specified environment. | [S6, S67] | 2 | 2% |
| Resource Utilization | It characterizes resources used, i.e., memory, cpu, disk and network usage. | [S105] | 1 | 1% |

**Table 13**
Categories of concerned application domains and related studies.

| Application domain | Primary study | # | % |
|---|---|---|---|
| Service-based application | General [S26, S37, S58, S64, S74, S81, S95, S97, S98], automated teller machine [S73, S89, S102, S103], course selection system [S14], Get Stream [S101], happy hour organizer [S54], internet banking system [S76], message sending system [S96], museum assistance system [S40], news provider [S1, S70], online training system [S3], online bookstore [S33], online course registration system [S66], online printer [S5], online shopping system [S50, S57, S60, S108], online train ticket system [S63], supply chain system [S6], travel reservation system [S20, S80], travel companion [S15], weather forecast system [S75] | 35 | 32% |
| Ambient assisted living system | [S16, S18, S22, S34-S36, S42, S43, S48, S49, S59] | 11 | 10% |
| MetaSocket | [S28, S30, S31, S 38, S91, S92, S100] | 7 | 6% |
| Ambulance dispatch system | [S44, S45, S51, S71, S106, S107] | 6 | 6% |
| Flood warning system | [S8, S17, S21, S24, S41, S52] | 6 | 6% |
| Mobile phone application | [S13, S69, S72, S85, S88, S93] | 6 | 6% |
| Remote data mirroring | [S68, S78, S79, S82, S83, S86] | 6 | 6% |
| Robot system | [S10, S56, S62, S84, S87, S105] | 6 | 6% |
| Intelligent Transport System | [S4, S94, S99, S104] | 4 | 4% |
| Washing machine system | [S2, S9, S25] | 3 | 3% |
| Adaptive programs | [S29, S47] | 2 | 2% |
| Conference management system | [S19, S23] | 2 | 2% |
| E-health application | [S77, S109] | 2 | 2% |
| Others | Dinner planner [S7], adaptive manager [S11], fire alert system [S12], rack server system [S27], production cell [S32], IDS project [S39], e-mobility [S46], load balancer system [S53], Cleaner agent [S55], order & delivery app [S61], image filter [S65], video surveillance system [S67], news application [S70], agent-based system [S90] | 13 | 12% |

*4.2.4. RQ2.4: What application domains were used for illustration?*

This question can be answered by investigating the illustration section of each primary study. The classification of the used systems or applications is provided in Table 13. It presents that service-based applications account for 32% (35/109) of the primary studies, followed by ambient assisted living system, MetaSocker, ambulance dispatch system, etc. Readers can find more details of each application in corresponding primary studies.

*4.3. RQ3: How well were the selected studies conveyed?*

This question is designed for addressing the quality of primary studies and comparing the quality between different modeling methods and different requirements activities. According to the quality assessment checklist, we evaluated the primary studies over each assessment questions and provide the detailed scores in [47]. The distribution of different scores is presented in Fig. 10. It depicts that the primary studies are scored between 11.0 and 17.0. Among these studies, 79% (86/109) are of good quality, i.e., scored between 15.0 and 17.0, because their qualities are highly assured by the selection criteria (Table 4). Only 4 studies are scored below 13.0.

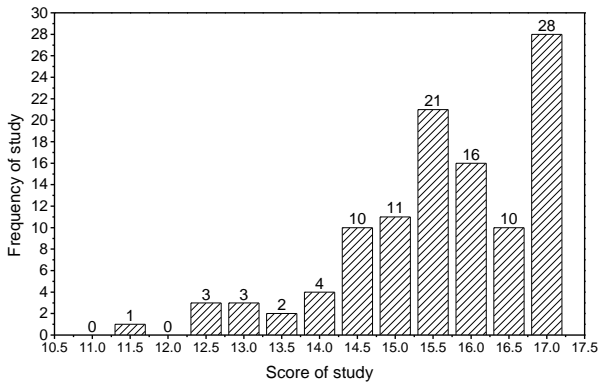

**Fig. 10.** Research distribution over scores.

*4.3.1. RQ3.1: How rigorously were the modeling methods applied and evaluated?*

The scores of primary studies can be aggregated and presented according to different modeling methods with descriptive statistics (Fig. 11). The mean score of each method is provided under the method name. For enterprise models (Fig. 11 (a)), the figure shows that PF and Tropos are better conveyed than others. However, PF is only involved in one study, so the sample size was too small to be statistically significant. For goal-oriented methods, though Tropos is better delivered than KAOS and i*, the scores present the characteristic of polarization. Most behavior models (Fig. 11 (b)) and logic models (Fig. 11 (c)) are assigned with high scores, because they provide avenues to rigorous model checking.

DNN, GA, GP and UT are highly qualified modeling methods (Fig. 11 (d)), since they are naturally mathematical models which can be computed with rigorous algorithms. Fig. 11 (e) presents the scores according to model categories. Since the sample sizes of enterprise models, behavior models and logic models are relatively large, their results are more convincing. The curve beside the box refers to the approximate normal distribution curve according to the scores. It depicts that behavior models are more rigorously conveyed, followed by enterprise models and logic models.

*4.3.2. RQ3.2: How much detail were the requirements activities presented with?*

The scores can also be aggregated according to the involved requirements activities (Fig. 12). For activities at requirements time (Fig. 12 (a)), specifying requirements is more highly scored than others, because the specification languages possess rigorous grammar. Most studies addressing adaptation mechanism and modeling requirements are assigned with scores below 17.0, because these two activities are more conceptual. Though the results of activities at design time and development time are also not statistically significant, we still provide the distribution of the scores for exhaustive description (Fig. 12 (b)). For activities at runtime (Fig. 12 (c)), most studies are scored between 15.0 and 17.0, because the illustrations of these studies are more detailed and the evaluations of the proposed approach are more explicit. According to Fig.12 (d), the scores of studies focusing on requirements time are more likely to be subject to the normal distribution. Besides, the activities at runtime were more detailed discussed than those at requirements time, design time and development time.

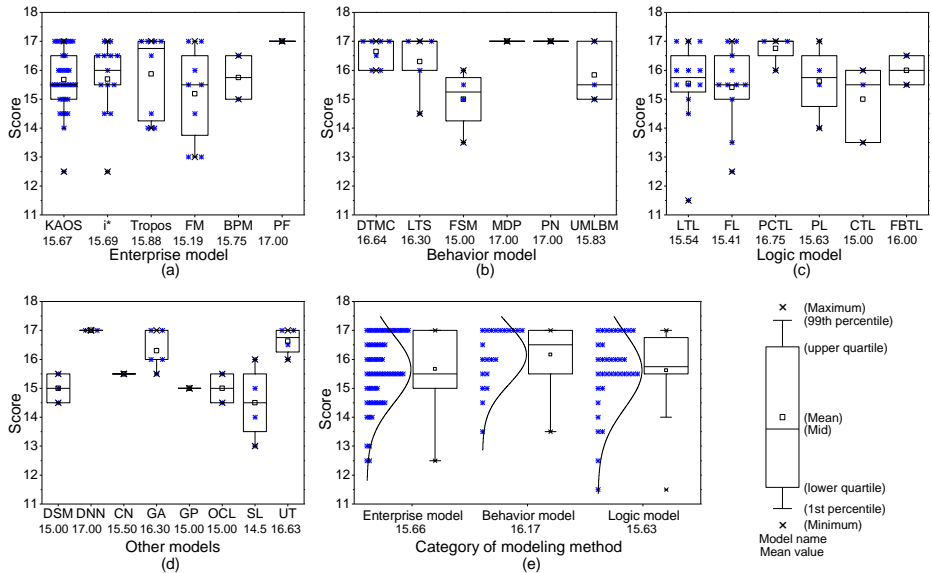

**Fig. 11.** Distribution of scores over modeling methods.

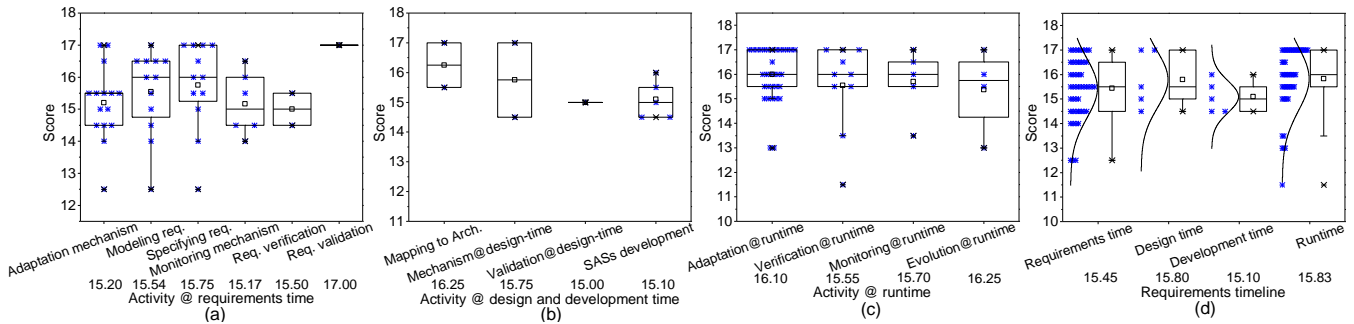

**Fig. 12.** Distribution of scores over requirements activities.

## 5. Discussions and recommendations

In the previous section, we elaborated the synthesized results, principal findings and basic discussion for each research question. This section provides further discussion as complements by highlighting advantages, identifying more research gaps and providing relevant recommendations.

### 5.1. Modeling methods and modeling dimensions

In [32], the authors proposed a classification of modeling dimensions and optional degrees for self-adaptive software systems which is depicted in Fig. 13. Each modeling dimension describes a particular facet of the system that is relevant to self-adaptation. We do not intend to exhaust all the definitions of each degree. Readers can find more details of these dimensions by referring to [32]. This subsection presents how well each dimension is studied in the literature and identifies research gaps in these dimensions. By considering these dimensions during modeling requirements and systems, the produced models will be more powerful.

#### 5.1.1. Challenges in goals dimensions

Goals refer to the objectives a system should achieve. *Evolution* dimension identifies whether the goals can change within the system lifecycle. It is known that adaptation deals with the changes in both systems (internal) and environment (external). From the perspective of computational reflection [57], the requirements of SASs can be represented with meta-level structures while the implementation can be represented with base-level structures. Therefore, the internal changes can occur to both meta-level elements, e.g., requirements and tasks, and base-level elements, e.g., components. According to our investigation, we find that most studies concentrate on task changes and environmental changes while only a few of studies concern the requirements changes. When requirements change at runtime, the pre-defined adaptation algorithms may lose effectiveness. Thus, the system should evolve from the old meta-level structure to the new one. To this end, both models@runtime methods and evolutionary computation methods can support the evolution process. *Flexibility* dimension captures whether the goals are expressed flexibly. It is related to the level of uncertainty associated with goal specifications. PCTL based specifications [S95-S97] and FL based specifications, e.g., FLAGS [S9] or RELAX [S34-S36], can support this dimension with probabilities and possibilities. *Duration* dimension refers to the validity of a goal through the system lifetime, i.e., temporary or persistent. It addresses the period of time when a goal is valid. Few existing studies consider this dimension. *Multiplicity* dimension characterizes how many goals there are. This dimension is involved in various kinds of modeling methods, e.g., goal models and feature models. When multi-goals are considered, the adaptation decisions are always made through trade-off among these goals, e.g., [S76-S79]. *Dependency* dimension captures how goals are related to each other. This dimension is widely conveyed in requirements models based on i* framework and Tropos method.

The results reveal the research gaps in the *flexibility* dimension and *duration* dimension. Utilizing linguistic variables and linguistic terms of fuzzy theory [58] can be a promising way to represent uncertainty and reasoning with uncertainty. Nevertheless, this kind of representation relies on the engineers' domain knowledge. To model the duration dimension, we suggest considering the Markov Renewal Process for describing the duration of the goal changes.

#### 5.1.2. Challenges in change dimensions

Change identifies the cause of adaptation. *Source* dimension characterizes the sources of changes. In RE literature, the most commonly considered sources are goals, system tasks, features and contexts. *Type* dimension refers to the nature of changes. According to the definition given by [32], functional changes refer to the changes needed for achieving the changed purpose of the system, while non-functional changes refer to changes needed for improving NFR, such as reliability and performance. How to adjust the system to satisfy FR is widely discussed in the literature. However, approaches for improving NFR still lack study. *Frequency* dimension captures how often a particular change occurs. This dimension is not considered in any primary studies. *Anticipation* dimension identifies whether a change can be anticipated. Foreseen changes refer to that we know when the changes occur during operation. To tackle this, the system just needs to take countermeasures at certain times. Foreseeable changes refer to that we know the changes will occur, but we do not know when they occur. To deal with this, adaptation mechanisms should be built for monitoring the changes, diagnosing impacts and making relevant decisions. Unforeseen changes identify the changes that cannot be anticipated. The system does not know what to monitor and the designed adaptation mechanisms will lose effectiveness. These changes are less concerned in the literature.

Research gaps lie in several aspects of the *change* dimension. To represent the satisfaction of NFR, we suggest using the notion of *utility* which can be calculated with utility functions. The improvement of NFR can be considered as the increase of the utilities of NFR. System configurations can be chosen with the objective function of optimizing the overall utility. Frequency of changes can also be modeled with a Markov Renewal Process. Besides, we recommend using the frequency of changes to calculate the overhead and test the responsiveness of the adaptation, for illustrating the efficiency of adaptation mechanisms. For coping with unforeseen changes, evolution of models at runtime can be a promising solution.

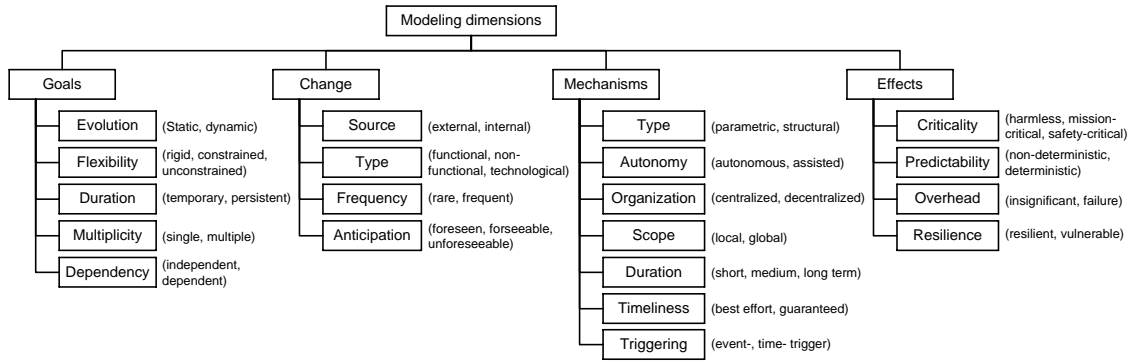

**Fig. 13.** Classification of modeling dimensions and optional degrees in [32].

*5.1.3. Challenges in mechanisms dimensions*

This set of dimensions captures the system reaction towards changes. *Type* dimension identifies whether the adaptation is related to the system parameters or system structures. These two types of adaptation have been described in [59] as parametric adaptation and structural or compositional adaptation. Most adaptation mechanisms focus on structural adaptation, such as adaptation mechanisms based on goal models [S59] and feature models [S67]. Parametric adaptation is discussed in [S23, S98] and [60]. *Autonomy* dimension characterizes the outside intervention during adaptation. All the primary studies focus on autonomous adaptation. *Organization* dimension captures whether the adaptation is done by a single component or distributed components. Most work utilizes the idea of centralized control, which is implemented by a central controller, such as the MAPE-based autonomic manager. Though a case study of decentralized systems is provided in [S94], the authors did not present how to build a decentralized control mechanism. *Scope* dimension characterizes whether adaptation is localized or globalized. This dimension is not described directly in the primary studies, but the proposed adaptation mechanisms imply that the adaptation is localized. *Duration* dimension refers to how long the adaptation lasts. According to the evaluations of the proposed approaches to adaptation and decision making at runtime, the duration of adaptation is short-termed. *Timeliness* dimension describes whether the time period for performing self-adaptation can be guaranteed. This dimension is closely related to the *frequency* dimension. When the change occurs frequently, the timeliness should be guaranteed with best effort. Several studies discussed time-consuming when the adaptation problem gets more complex or the model gets larger. The evaluations imply that the timeliness is guaranteed. *Triggering* dimension characterizes whether the change that initiates adaptation is event-trigger or time-trigger. All the primary studies consider adaptation is triggered by event, i.e., requirements violations.

Research gaps are involved in parametric adaptation, decentralized adaptation and time-triggered adaptation. To provide parametric adaptation, we need to conduct the system identification process and specify the mathematical relationships between changes and effected requirements or system tasks. Decentralized control is a flexible way to achieve adaptation, when the system is composed of distributed parts. The centralized control may be limited by the physical positions of these parts and environmental resources, e.g., bandwidth. Therefore, endowing the distributed parts with the self-adaptability is a way to overcome these drawbacks. To this end, adaptation manager and adaptation algorithms should be built into each distributed part for achieving local control and adaptation. For event-triggered adaptation, the system should continuously monitor the identified event at runtime. When the event is of low frequency, the monitoring process would increase the overhead. A promising way to deal with this situation is using time-triggered adaptation. However, the precondition is that we should know the general time or time interval when changes will occur.

*5.1.4. Challenges in effects dimensions*

Effects identify the impact of adaptation upon the system. *Criticality* dimension characterizes the impact upon the system in case of adaptation fails. Most primary studies focus on the mission-critical adaptation. In Table 13, the primary studies utilizing ambient assisted living systems, flood warning systems and e-health applications imply the consideration of safety-criticality. *Predictability* dimension captures whether the consequences of adaptation can be predictable. The deterministic predictability of adaptation is only provided in [S80, S81]. *Overhead* dimension describes the negative impact of adaptation upon system's performance. Though some studies illustrate the response time of the adaptation, they do not discuss the system failure caused by high-overhead. *Resilience* dimension characterizes whether the system can persistently achieve its objective when facing changes. Resilience is well supported by the FL-based approaches, which can be found by referring to Table 9.

Both *predictability* dimension and *overhead* dimension need to be further studied. The techniques can be various according to the proposed models and illustrating examples. Besides, this set of dimensions is closely related to the assurance of adaptation.

*5.2. Requirements activities and assurance dimensions*

This subsection provides several research gaps and recommendations for the identified requirements activities, and discusses what should be concerned in these activities to support the assurance of self-adaptation.

*5.2.1. Challenges at requirements time*

According to the extracted data, we find that researchers were prone to employ feedback control loop to engineer SASs, particularly, the MAPE-K loop. Feedback control loop adjusts the input(s) according to the measured error and maintains the output(s) sufficiently closed to what is desired. Therefore, it can be viewed as a retroactive control mechanism. However, other control paradigms can also provide their own advantages. Feedforward control loop measures the disturbances and adjusts the control input to reduce the impact of the disturbance on the system output. Thus, it is considered as a proactive control mechanism. We suggest considering feedforward-feedback control mechanism has the advantage of both control schemes. First, it can tune system behavior based on the measured disturbances at runtime. Second, when deviations exist between the measured output and desired output, it can correct the behavior accordingly. Besides, fuzzy control is also an effective control scheme. Fuzzy control is a practical alternative for achieving high-performance control on nonlinear time-variant system since it provides a convenient method for constructing nonlinear controllers using heuristic rules. Engineers incorporate these rules into a fuzzy controller that emulates the decision-making process of the human [61].

When modeling requirements, researchers have proposed how to model contexts and how to incorporate contextual information into requirements model [62, 63]. However, environmental behavior was not thoroughly discussed in the literature. If we consider the environment as an entity, its behavior can also be modeled by a set of contextual states and transitions. Contextual uncertainty can be represented as probabilistic transitions. By integrating environmental behavior with system behavior, we can produce more powerful models, such as Hidden Markov Model (HMM). For monitoring requirements, both qualitative methods (e.g., [S43, S102]) and quantitative methods (e.g., [S45, S101, S104]) were proposed. However, few studies focus on how to utilize contextual information and integrate it with requirements.

*5.2.2. Challenges at design time*

At design time, how to map requirements models to architectural models has been conveyed in the literature [S49, S50]. We suggest building adaptation mechanisms with bidirectional transformations between requirements models and architectural models. In this way, the system can respond to internal changes in both requirements and components. When components failure or noise occurs, it will be mapped to requirements models. Then, adaptation decisions can be made by reasoning over requirements mod-els. After mapping the decisions to components models, the system can operate with new configurations.

*5.2.3. Challenges at Runtime*

At runtime, more work needs to be conducted to support requirements verification with contextual information. For example, the system behavior is modeled with LTS, while the transitions are annotated with contextual information. Besides, other models except for KAOS need to be explored to support models@runtime and system evolution, e.g., i*, BPM and DTMC. Studies that focus on runtime adaptation need to illustrate the assurance of adaptation. Researchers in [31] proposed 17 assurance criteria for self-adaptive systems. From the perspective of *effectiveness*, *effects* and *consumption*, a set of 8 criteria and three extended aspects should be considered as the assurance dimensions for runtime adaptation (Table 14). Effectiveness captures the validity of the adaptation. Effects characterize the impact of adaptation upon the system. Consumption is concerned with the consumed time and resources. By illustrating these dimensions, researchers can justify the trustworthiness of their approaches and compare their results with others'.

**Table 14**
Assurance dimensions for self-adaptation.

| Category | Dimension | Degree | Description |
| --- | --- | --- | --- |
| Effectiveness | Confidence | none, low to high | Whether the adaptation can satisfy all the requirements |
| | Feasibility | never, sometimes to always | Whether the adaptation can be carried out all the time |
| | Coverage | none, small to large | Whether all the changes can be solved through adaptation and whether the adaptation can lead to all the possible solutions |
| Effects | Fitness | unacceptable, acceptable, optimal | How good the adaptation results are |
| | Determinism | unrepeatable to repeatable | Whether same conditions lead to same adaptation |
| | Resilient | resilient to vulnerable | After adaptation, whether the system needs to readjust to all the changes |
| | Predictable | non-deterministic to deterministic | Whether the consequences of adaptation can be predictable |
| | Overhead | Insignificant to failure | The impact of adaptation upon system performance |
| Consumption | Duration | short, medium, long term | The time used during adaptation process |
| | Sustainability | small to large | The resources used during adaptation process |
| | Timeliness | best-effort to guaranteed | Whether the time period for performing adaptation can be guaranteed |

## 6. Threats to validity

To mitigate systematic errors in design and conduction of the literature review, we established a rigorous protocol in advance which was reviewed by both internal and external reviewers. This section discusses possible threats to the validity of our literature review.

*6.1. Threats to search sources*

Threats to search sources might arise during choosing publication venues and digital libraries. The defined publication venues for manual search may not cover all the publications of the literature. We mitigated this threat by conducting the automated search and the "snowballing" search. We limited the manual search based on publication venues using the ERA ranking and the H-index. We believe that by including in top journals, conferences and symposiums, the quality of the results of our SLR can be improved a lot. To guarantee thoroughly retrieved results, we identified eight popular digital libraries that almost cover all the publications of the literature.

*6.2. Threats to search strategies*

Threats to search strategies might arise during defining search strings and searching with the strings. To avoid researchers' bias, we adopt QGS to objectively define search strings. The QGS was established based on the integrated manual search results of the two reviewers. By mining the meta-data (title-abstract-keywords) of the papers in the QGS, search terms were derived according to the term frequency and inverse document frequency while search strings were generated based upon Jaccard's similarity coefficient.

Since different digital libraries provide different capabilities to search for publications, we conducted a pilot study in each digital libraries to check the validity and feasibility of the search strings. To illustrate the efficiency of the search strings, we calculated the *quasi-sensitivity*. The results confirmed the defined strings to be optimum in our SLR (Fig. 4). Though the search strings can be broad, it is possible that they are not able to capture some studies. It can be mitigated through a "snowballing" search. Moreover, the time scope of automated search is from 2003 to 2013. There possibly exist several relevant studies before 2003. However, the selected results (Fig. 6) depict that no studies existed in 2003 and 2004, which may indicate the possibility is quite slim.

*6.3. Threats to selection strategies*

Threats to selection strategies might arise during defining selection criteria and selecting primary studies. The main threat to the validity of selection criteria is that we excluded editorial, abstract, keynote, poster and short paper within 6 pages or less (C6), opinion papers and position papers which have no detailed description of modeling methods (C7), and all the secondary studies (C8). Actually, we did not intend to exhaustively list all the publications in the literature. The objective of this study is to primarily investigate the support of modeling methods for requirements activities. Therefore, we only focus on the studies which involve more detailed descriptions of modeling methods, requirements activities and demonstrations. Meanwhile, these highly qualified studies can improve the quality of our SLR. We also conducted a pilot study for ensuring the effectiveness of the selection criteria.

The main threat to the selection process is researchers' bias. To reduce such bias, we produced a well-defined protocol to guarantee the consistency in the selection of primary studies. Before selection, we conducted a pilot study to ensure the validity of the defined selection process. The selection process was conducted in parallel by two researchers and a cross-check was performed after each selection round. Besides, we calculated Kappa coefficient to certify the inter-rater agreement between the two researchers. When any disagreement between the researchers occurred, we conducted a joint meeting with other researchers and eliminated the disagreement through discussion.

*6.4. Threats to data extraction*

Threats might also arise during the data extraction process. First, we utilized the protocol to ensure the consistency of the extraction process for two researchers. The extraction results heavily rely upon their understanding of the content of primary studies, especially modeling methods and requirements activities.

To minimize the researchers' bias, we still leveraged Kappa coefficient, cross-check and joint meeting.

To avoid the bias in quality assessment, we designed ten assessment questions which almost cover all aspects of a paper. Different from the checklist introduced in the SLR guidelines [13], we designed a weighted quality assessment checklist. When we conducted a pilot study with the non-weighted assessment questions, we found that a study with relatively detailed context description may score higher than a study with relatively detailed modeling methods. Since this SLR focuses on modeling methods, this phenomenon is considered as a potential threat to the validity of quality assessment. To eliminate such threat, we assigned a weight to each question and the final score of a paper turns to be a weighted value.

Moreover, during extracting quality attributes and application domains, we found that some papers insufficiently describe the corresponding information. It can also be viewed as a potential threat to data extraction. Therefore, we had to infer certain pieces of information and discussed among researchers to verify the correctness.

## 7. Related work

This section presents and discusses some previous secondary studies relevant to self-adaptive systems. From this perspective, we identified three studies related to our work, which focus on control engineering approaches, formal methods and research subjects, respectively. The differences between these studies and our study are presented in Table 15.

Patikirikorala et al. [9] performed a systematic survey on applying control engineering approaches to the design of self-adaptive systems. They totally selected 161 papers from different publication venues with the time span from 2000 to 2011. The authors 1) built a classification model of the existing literature; 2) quantified the published research work on the various modeling, control schemes and validation techniques; and 3) analyzed the clustering of papers across categories of the classification model and identified research trends. However, we consider using different control schemes to engineer SASs from RE perspective. More specifically, we focus more on how to apply these control schemes to build effective adaptation mechanisms at requirements time.

Weyns et al. [10] conducted a well-designed systematic survey to identify what formal methods had been used in the SASs field and for what purposes these methods have been used. They collected 75 papers from 12 conferences and workshops and 4 journals with the time span from 2000 to 2011. They 1) presented research trends in applying formal methods; 2) identified the used modeling languages and property specification languages; and 3) categorized the concerned quality attributes, software systems and verified properties. Different from their work, we categorize the modeling methods according to the nature of usage. Apart from the advantage of systematization, our SLR contains (but is not limited to) investigating formal methods and verification activities. Besides, the synthesized results and the generated conclusions of our SLR are based on more relevant studies and more recently published work.

Weyns et al. [11] performed another literature study to summarize existing research related to engineering SASs and shed light on what the claimed benefits are of adaptation and to what extent evidence exists for these benefits. They selected 96 papers from the proceedings of SEAMS between 2006 and 2011 and the papers in the book of Dagstuhl Seminar 08031. By analyzing these papers, they categorized the research subjects, summarized the used application domains, identified their concrete focus and claimed benefits. Different from their work, we concentrate on the RE aspect of self-adaptive systems, particularly the requirements modeling and analysis aspect. Apart from the classification of modeling methods, we also provide a systematic classification of requirements activities which is useful for researchers and practitioners to better understand the literature. Besides, we present how the identified modeling methods can support each requirements activity and discuss the modeling dimensions and the assurance dimensions.

**Table 15**
Comparison between related SLRs.

| Related SLR | Literature | # of RQ | Time span | # of digital library | # of relevant venues | # of selected papers | Consider modeling methods | Consider requirements activities | Consider engineering SASs |
|---|---|---|---|---|---|---|---|---|---|
| Patikirikorala et al. [9] | Control engineering for SASs | 5 | 10 years | 4 | 12 | 161 | × | × | √ |
| Weyns et al. [10] | Formal method for SASs | 4 | 10 years | N/A | 16 | 75 | √ | × | × |
| Weyns et al. [11] | SASs | 5 | 6 years | N/A | 2 | 96 | × | × | × |
| Our SLR | RE of SASs | 9 | 10 years | 8 | 21 | 109 | √ | √ | √ |

## 8. Conclusions and future work

During the past decade, an increasing number of publications in the RE community indicates the growing interest in self-adaptive systems. In this article, we report a systematic literature review that investigates and summarizes the state of the art on requirements modeling and analysis for self-adaptive systems. More specifically, the aim of this SLR is to 1) present the publication distributions; 2) identify the used modeling methods, involved requirements activities, software quality attributes and application domains; 3) present what support these modeling methods can provide; 4) assess the quality of primary studies over modeling methods and requirements activities; and 5) shed light on research gaps and provide recommendations.

Our SLR is underpinned by a rigorously defined protocol, which has been reviewed by internal and external experts. By systematically analyzing the selected 109 papers, we gain a series of findings, implications and conclusions:

- Research groups in USA and UK lead more research work with fewer researchers.
- A classification of modeling methods is proposed (Table 6), which is composed of 8 categories and 29 sub-categories.
- The most widely used enterprise modeling method is KAOS method.
- A novel classification of requirements activities is provided based on requirements activity timelines (Table 11).
- The most frequently considered requirements activity is requirements-driven runtime adaptation.
- Not all quality attributes were studied and over half of the selected studies only concern adaptability.
- Changeability and time behavior were supported by more modeling methods than other quality attributes.
- Service-based applications are utilized by most studies as illustrating examples.

- Mobile phone applications were supported by more modeling methods than others.
- The mean values vary a lot from one modeling method or requirements activity to another.
- Studies utilizing behavior models and focusing on runtime requirements activities receive higher scores, because the approaches were more rigorously evaluated.

As complement, we discussed the relationships between modeling methods and modeling dimensions, and the relationships between requirements activities and assurance dimensions. Meanwhile, research gaps were identified for each modeling dimension and each requirements timeline.

This work will contribute to the understanding of requirements modeling and analysis for self-adaptive systems. Researchers and practitioners from both the RE field and the SASs field will benefit from this work.

- For junior researchers, this work can serve as a starting point and landscape for future research into the literature of RE for SASs.
- For senior researchers, this work can serve as a systematic summarization of the literature over the past decade and the identified research gaps may inspire them to produce more research results in the needed areas.
- For practitioners, various kinds of monitoring mechanisms and adaptation mechanisms can help them develop a new set of tools to monitor changes and detect requirements violations. Besides, they can also benefit from the identified application domains and develop concrete techniques and prototype systems.

From the review findings, we perceive that there exist lots of promising directions and interesting problems in SASs field for the RE community to discover and solve, e.g. building adaptation mechanism with self-adaptive control loops and dealing with the uncertainty caused by sensor error. The challenges in modeling dimensions and assurance dimensions are all need to be further considered. Meanwhile, more systematic studies are needed for seeking a better understanding of the literature. Future work could be carried out to review the assurance method for requirements-driven adaptation.

## Acknowledgements

<>We thank Prof. Barbara A. Kitchenham and her team at Keele University for reviewing our protocol and all the received advice. This research is supported by the National Natural Science Foundation of China under Grant Nos 61232015 and 91318301, and in part by the Natural Science Foundation of Guangxi Province under Grant No. 2012GXNSFCA053010.## Appendix A. Search strings

The table presents the final search strings clustered according to requirements activities.

| ID | Activities | Search strings |
|---|---|---|
| S1 | Modeling | DOMAINS[a] AND ("model requirements" OR "modeling requirements" OR "requirements modeling") |
| S2 | Specifying | DOMAINS AND ("specify requirements" OR "specifying requirements" OR "requirements specifying" OR "requirements specification") |
| S3 | Monitoring | DOMAINS AND ("monitor requirements" OR "monitoring requirements" OR "requirements monitoring") |
| S4 | Awareness | DOMAINS AND ("aware requirements" OR "requirements-aware" OR "requirements awareness" OR "requirements-awareness") |
| S5 | Diagnosing | DOMAINS AND ("diagnose requirements" OR "diagnosing requirements" OR "requirements diagnosing" OR "requirements diagnosis") |
| S6 | Detecting | DOMAINS AND ("detect requirements" OR "detecting requirements" OR "requirements detection") |
| S7 | Verifying | DOMAINS AND ("verify requirements" OR "verifying requirements" OR "requirements verifying" OR "requirements verification") |
| S8 | Adaptation | DOMAINS AND requirements AND (adaptation OR reconfiguration OR decision) |
| S9 | Evolution | DOMAINS AND ("evolution requirements" OR "requirements evolution") |

[a] **DOMAINS** denotes the string: ("self-adaptive systems" OR "dynamically adaptive systems" OR "self-adaptive software" OR "autonomic computing").

## Appendix B. Primary studies

[S1] L. Baresi, L. Pasquale, Adaptation goals for adaptive service-oriented architectures, in: P. Avgeriou, J. Grundy, J.G. Hall, P. Lago, I. Mistrík (Eds.), Relating Software Requirements and Architectures, Springer, 2011, pp. 161-181.

[S2] L. Baresi, L. Pasquale, Adaptive goals for self-adaptive service compositions, in: Proceedings IEEE International Conference on Web Services (ICWS), IEEE, 2010, pp. 353-360.

[S3] Y. Jian, L. Liu, J. Wang, E. Yu, Goal-driven adaptation of internetware, in: Proceedings of the Second Asia-Pacific Symposium on Internetware, ACM, 2010, pp. 1-9.

[S4] Y. Jian, T. Li, L. Liu, E. Yu, Goal-oriented requirements modelling for running systems, in: Proceedings of the First International Workshop on Requirements@Run.Time, IEEE, 2010, pp. 1-8.

[S5] Z. Yang, Z. Jin, Requirements modeling and system reconfiguration for self-adaptation of Internetware, in: Proceedings of the Fourth Asia-Pacific Symposium on Internetware, ACM, 2012, pp. 1-6.

[S6] Z. Ma, L. Liu, H. Yang, J. Mylopoulos, Adaptive service composition based on runtime requirements monitoring, in: Proceedings IEEE International Conference on Web Services (ICWS), IEEE, 2011, pp. 339-346.

[S7] L. Baresi, L. Pasquale, Live goals for adaptive service compositions, in: Proceedings of the ICSE Workshop on Software Engineering for Adaptive and Self-Managing Systems (SEAMS), IEEE, 2010, pp. 114-123.

[S8] H.J. Goldsby, B.H.C. Cheng, Automatically generating behavioral models of adaptive systems to address uncertainty, in: Proceedings of the 11th International Conference on Model Driven Engineering Languages and Systems (MODELS), Springer, 2008, pp. 568-583.

[S9] L. Baresi, L. Pasquale, P. Spoletini, Fuzzy goals for requirements-driven adaptation, in: Proceedings of the 18th IEEE International Conference on Requirements Engineering (RE), IEEE, 2010, pp. 125-134.

[S10] N. Bencomo, K. Welsh, P. Sawyer, J. Whittle, Self-explanation in adaptive systems, in: Proceedings of the 17th International Conference on Engineering of Complex Computer Systems, IEEE, 2012, pp. 157-166.

[S11] R. Ali, C. Solis, I. Omoronyia, M. Salehie, B. Nuseibeh, Social adaptation at runtime, in: L. Maciaszek, J. Filipe (Eds.), Evaluation of Novel Approaches to Software Engineering, Springer Berlin Heidelberg, 2013, pp. 110-127.

[S12] E. Santos, J. Pimentel, D. Dermeval, J. Castro, O. Pastor, Using NFR and context to deal with adaptability in business process models, in: Proceedings of the Second International Workshop on Requirements@Run.Time, IEEE, 2011, pp. 43-50.

[S13] M. Salifu, Y. Yijun, B. Nuseibeh, Specifying monitoring and switching problems in context, in: Proceedings of the 15th International Conference on Requirements Engineering (RE), IEEE, 2007, pp. 211-220.

## Appendix C. Quality assessment checklist

Each of the 10 assessment questions has a weight and 3 optional answers.

| Assessment question | Optional answer | Score | weight |
|---|---|---|---|
| **A1:** How clearly is the research questions/issue of study stated? (Problem statement) | Explicitly (clearly defined in the introduction) | 1 | 1 |
| | Generally (cannot figure out directly) | 0.5 | |
| | Vaguely (hard to figure out) | 0 | |
| **A2:** How adequately is the context in which the research is carried out described? (Context description) | Explicitly (discuss the limitation of existing studies) | 1 | 1 |
| | Generally (only describe related work) | 0.5 | |
| | Vaguely (hard to figure out) | 0 | |
| **A3:** How clearly is the modeling method conveyed? (Modeling methods) | Operationally (modeling process can be repeated by steps) | 1 | 2 |
| | Explicitly (hard to repeat the modeling process) | 0.5 | |
| | Briefly (understandable but not repeatable) | 0 | |
| **A4:** How clearly is the requirements activities conveyed? (Requirements activities) | Operationally (Requirements activities can be repeated by steps) | 1 | 2 |
| | Explicitly (hard to repeat requirements activities) | 0.5 | |
| | Briefly (Understandable but not repeatable) | 0 | |
| **A5:** How detailedly is the proposed approach evaluated? (Evaluation) | Detailedly (with detailed description) | 1 | 3 |
| | Briefly (with basic steps and results) | 0.5 | |
| | Illustratively (with "toy examples") | 0 | |
| **A6:** How well do the evaluation results address its research questions? (Validity) | Entirely (all research questions are answered) | 1 | 2 |
| | Partially (parts of questions are answered) | 0.5 | |
| | Vaguely (questions seem not to be answered) | 0 | |
| **A7**: Do the researchers explain any problems when applying their approach? (Threats to validity) | Explicitly (discuss threats and provide suggestions) | 1 | 1 |
| | Generally (only describe threats) | 0.5 | |
| | None (no description) | 0 | |
| **A8:** How well are results and findings are discussed? (Discussion) | Explicitly (discuss advantages and limitations) | 1 | 2 |
| | Generally (only describe the general results) | 0.5 | |
| | Vaguely (no clearly description) | 0 | |
| **A9:** How well are the approach and results compared with related work? (Related work) | Explicitly (with detailed comparison) | 1 | 2 |
| | Generally (with brief comparison) | 0.5 | |
| | Vaguely (only describe related work) | 0 | |
| **A10**: How does the approach contribute to the literature? (Contributions) | A new approach to a new research question | 1 | 2 |
| | A new approach to an existing research question | 0.5 | |
| | An existing approach to an existing research question | 0 | |

**Figure captions**

Fig.1. Systematic literature review process.
Fig.2. Mechanism underpinning the search process (adapted from [42]).
Fig.3. Clustering result of high frequency terms.
Fig.4. Quasi-sensitivity for each digital library and the total quasi-sensitivity.
Fig.5. Details of the search and selection process.
Fig.6. Research distribution over the last decade.
Fig.7. Research distribution over research groups.
Fig.8. Research distribution over countries.
Fig.9. Research distribution over continents.
Fig.10. Research distribution over scores.
Fig.11. Distribution of scores over modeling methods.
Fig.12. Distribution of scores over requirements activities.
Fig.13. Classification of modeling dimensions and optional degrees in [32].

**Table captions**

Table 1 Research questions.
Table 2 Conferences and symposiums.
Table 3 Journals and books.
Table 4 Selection criteria.
Table 5 Frequency of papers in selected publication venues.
Table 6 Categories and descriptions of modeling methods.
Table 7 Enterprise models and related studies.
Table 8 Behavior models and related studies.
Table 9 Logic models and related studies.
Table 10 Domain models, network models, evolutionary computation models, language models, utility models and related studies.
Table 11 Categories of requirements activities and related studies.
Table 12 Descriptions of concerned software quality attributes and related studies.
Table 13 Categories of concerned application domains and related studies.
Table 14 Assurance dimensions for self-adaptation.
Table 15 Comparison between related SLRs.